\documentclass[]{raa}           
\usepackage{graphicx,times}
\usepackage{natbib}
\usepackage{amssymb,amsmath}
\bibpunct{(}{)}{;}{a}{}{,}

\begin{document}

  \title{Analysis of interval constants of Shoushili-affiliated calendars}

 \volnopage{ {\bf 2012} Vol.\ {\bf X} No. {\bf XX}, 000--000}
   \setcounter{page}{1}

   \author{Byeong-Hee Mihn\inst{1,2}, Ki-Won Lee\inst{3}, Young Sook Ahn\inst{1}
   }

   \institute{Advanced Astronomy and Space Science Division,
Korea Astronomy and Space Science Institute, Daejeon 305-348, Korea \\
	\and
	 Department of Astronomy and Space Science, 
         Chungbuk National University, Cheongju 361-763, Korea\\
        \and
             Institute of Liberal Education, Catholic University of Daegu, 
Gyeongsan 712-702, Korea; {\it leekw@cu.ac.kr}\\
\vs \no
   {\small Received 2013 June 12; accepted 2013 July 27}
}

\abstract{
We study the interval constants that are related to the motions of
the Sun and the Moon, i.e., the Qi, Intercalation, Revolution, and Crossing
interval constants, in calendars affiliated with the Shoushi calendar
(Shoushili), such as Datongli and Chiljeongsannaepyeon. 
It is known that those interval constants were
newly introduced in Shoushili and revised afterward, except for
the Qi interval constant, and the revised values were  
adopted in the Shoushili-affiliated calendars. 
In this paper, 
we investigate first the accuracy of the interval constants and then 
the accuracy of the Shoushili-affilated calendars
in terms of the interval constants by comparing 
the times of the new moon and the solar eclipse maximum 
calculated by each calendar with
modern calculations. 
During our study, 
we found that the Qi and Intercalation interval constants used in 
the early Shoushili were well determined, whereas 
the Revolution and Crossing interval constants 
were relatively poorly measured . 
We also found that the interval constants used by the early 
Shoushili were better than those of the later one, and hence than those
of Datongli and Chiljeongsannaepyeon. 
On the other hand, we found that the early Shoushili is, in general,
a worse calendar than Datongli for use in China but a better one than
Chiljeongsannaepyeon for use in Korea in terms of the
new moon and the solar eclipse times, at least for the period 1281 -- 1644.
Finally, we verified that the sunrise and sunset times recorded in 
Shoushili-Li-Cheng and Mingshi are 
those at Beijing and Nanjing, respectively.
}

   \authorrunning{B.-H. Mihn et al. }            
   \titlerunning{Interval constants of Shoushili-affiliated calendars}
   \maketitle

%
\section{Introduction}           
\label{sect:intro}


The Shoushi calendar (Shoushili, hereinafter) was developed
by Shoujing Guo and his colleagues from the Yuan dynasty \citep{bo1997}. 
It is known as one of most the famous calendars in Chinese 
history \citep{needham1959}. Its affiliated calendars, the Datongli of the Ming 
dynasty and the Chiljeongsannaepyeon (Naepyeon, hereinafter) of the Joseon 
dynasty, are extremely similar. Compared to
previous Chinese calendars, the most distinguishing characteristic of 
the Shoushili is the abolition of the use of the Super Epoch, an ancient 
epoch where the starting points of all interval constants are the 
same \citep{sivin2009}. 
Instead, Shoushili adopted the winter solstice of 
1280 (i.e., 1280 December 14.06 in Julian date) 
as the epoch and introduced seven 
interval constants from observations in order to reduce the large 
number of accumulated days in the calculations arising from the use 
of the Super Epoch.  Of the seven interval constants, four are related 
to the motions of the Sun 
and the Moon: the Qi, Ren (Intercalation), Zhuan (Revolution), and Jiao
(Crossing) interval constants. 
According to the records of Mingshi (History of the Ming Dynasty), the
early values of these interval constants,
except for the Qi interval constant,
were revised later. Currently, it is known that Datongli and
Naepyeon adopted the revised values of the early Shoushili;
hence, both calendars are essentially
identical to the later Shoushili. The main purpose of this study is
to evaluate the effect of the values of these four interval constants
on the accuracy of calendar calculations. In this paper, therefore,
``Shoushili" refers to the early one 
unless stated ohterwise.

In this study, we first estimate the accuracy of 
the values of the Qi, Intercalation, Revolution, 
and Crossing interval constants of the Shoushili-affiliated calendars.
We then investigate the accuracy of each calendar in terms of the
interval constants by comparing the timings of the new moon and the solar 
eclipse maximum calculated by each calendar with modern calculations. 
Because sunrise and sunset times are needed to calculate 
the timings of the solar eclipse maximum 
in the Shoushili-affiliated calendars, we verify 
those times presented in calendar books as well.

This paper is composed as follows. 
In section~\ref{calendars}, we briefly introduce 
the Shoushili-affiliated calendars and modern astronomical calculations. 
In section~\ref{results}, we present the results on
the analysis of the four interval constants, 
the new moon times, the sunrise and sunset times, and the solar
eclipse maximum times. 
Finally, we summarize our findings in section~\ref{summary}.

\section{The Shoushili-affiliated calendars}
\label{calendars}

\subsection{Shoushili}
\label{shoushili}

In the chapter related to calendars in the Yuanshi 
(History of the Yuan Dynasty), 
Shoushili is described as covering two parts: Shoushili-Yi (Discussion) 
and Shoushili-Jing (Method). It is known that the calendar was used in the 
Goryeo dynasty of Korea since the reign of King Chungseon (1308 -- 1313)
\citep{jeon1974}.
For this reason, Shoushili is also recorded in the chapter related to 
calendars in the Goryeosa (History of the Goryeo Dynasty). 
However, the Goryeosa does not include the Shoushili-Yi. 
Conversely, the Yuanshi contains 
no Li-Cheng (Ready-reference Astronomical Tables), which is presented in 
the Goryeosa. In addition, both historical books have no tables on the sunrise 
and sunset times.

A book entitled Shoushili-Li-Cheng (Ready-reference Astronomical
Tables for Shoushili) is preserved in the 
Kyuganggak Institute for Korean Study (Kyuganggak, hereinafter) in Korea. 
This book contains various tables for the use of calendar calculations 
by Shoushili along with the times of sunrise and sunset 
times \cite[for further details, see][]{lee1998}. We have to refer to
three books (i.e., Yuanshi, Goryeosa, and Shoushili-Li-Cheng) to 
completely understand Shoushili. Kyuganggak also possesses a book 
called Shoushili-Jie-Fa-Li-Cheng (Expeditious Ready-reference Astronomical
Tables for Shoushili). According to the preface of the Kyuganggak edition, 
this book was brought from China by Bo Gang (an astronomer in the Goryeo 
court). It was printed in 1346, and reprinted in 1444, around the publication 
year of Naepyeon.

\subsection{Datongli}
\label{datongli}

There are two versions of Datongli in the Ming dynasty; Wushen-Datongli 
(Datongli of the Wushen Year) made by Ji Liu in 1368, and 
Datong-Lifa-Tonggui (Comprehensive Guide to the Calendrical
Method by Datongli) made by Tong Yuan in 1384. 
Although the epochs of the former and the latter calendars are 
the winter solstices of 1280 and 1383, respectively, both are
based on the Shoushili \citep{lee1996}. 
In the chapter related to calendars 
in the Mingshi, Datongli is described 
as having three parts: the first is on the origin of the techniques, 
the second on Li-Cheng including the tables on the timings of 
sunrise and sunset, and the third on calendar methods.

It is widely known that Datongli was used in the Goryeo dynasty since 
1370, although this fact is arguable \cite[refer to][]{lee2010}. 
Unlike Shoushili, Datongli is not included in the chapter related to
calendars in the Goryeosa. Instead, a series with a name similar
to Datong-Lifa-Tonggui of Tong Yuan, for example, Datong-Liri-Tonggui 
(Comprehensive Guide to the Calendar Day by Datongli), is
preserved in the Kyuganggak. According to the work of \cite{lee1988}, 
the series (Tonggui series, hereinafter) is related to the Datongli and was 
published around 1444 and its main purpose was being
a reference during the compilation of the Naepyeon.

\subsection{Naepyeon}
\label{naepyeon}

King Sejong of the Joseon dynasty ordered In-Ji Jeong et al. to compile 
the Naepyeon in 1433. Although the data on when the compilation was completed
is not clear,
the oldest extant version is the one published in 1444 by Sun-Ji Yi and
Dam Kim. In terms of the contents, each chapter of the Naepyeon corresponds to 
the Tonggui series and the timings of sunrise and sunset are contained in 
the Taeum (Moon) chapter. In addition, these times are different in the 
Shoushili-affiliated calendars. Because the compilation of the Naepyeon was 
considered to be one of the greatest works of King Sejong, this book is 
also appended in his Veritable Record, unlike Veritable Records of
other king's \citep{lee2008a}.

A book entitled Jeongmyoyeon-Gyeosik-Garyeong
(Example Supplement for the Calculations of the Solar and Lunar Eclipses 
Occurred in 1447; shortly Garyeong), which is related to the
Naepyeon,  also remains in the Kyuganggak. 
This book contains modified interval constants for the epoch of 1442 
and is valuable for step-by-step calculations of not only
the solar eclipse but also the new moon 
by the Naepyeon, and hence by the Shoushili or Datongli. 
In this study, we refer to Garyeong to calculate the new moon and
solar eclipse maximum times by the Shoushili-affiliated calendars, and
we refer to the work of Lee (1988), which introduced
the differences among the calendars.

In Table~\ref{tab1}, we summarize the values of the four interval constants 
in each calendar. All dates are given in the Julian 
calendar and all values of interval constants are in units of Part; 
one day is 10,000 Parts. 
The Qi interval constant is the interval between the epoch 
and the midnight of the first day in a sexagenary cycle on counting backwards 
from the epoch. This value is same in all Shoushili-affiliated
calendars. Intercalation interval constant is the interval from the 
epoch to the `mean' new moon of the month belonging to the epoch. 
The time of a new moon 
is determined by correcting the slowness or fastness of the solar and 
lunar motions to the `mean' new moon time, which is obtained by accumulating 
the Intercalation interval constant. Revolution and Crossing interval 
constants are the lengths between the epoch, and the times of lunar perigee
and descending node passage, respectively \cite[see also][]{sivin2009}.

\begin{table}
\bc
\begin{minipage}[]{150mm}
\caption[]{Summary of the four interval constants adopted in 
the Shoushili-affiliated calendars\label{tab1}.}\end{minipage}
\setlength{\tabcolsep}{2.5pt}
\small
 \begin{tabular}{lccccc}
  \hline\noalign{\smallskip}
Interval  & \multicolumn{2}{c}{Shoushili} & \multicolumn{2}{c}{Datongli} &
Naepyeon\\
\cline{2-5}
Constants & Yuanshi & Goryeosa & Mingshi$^1$ & Tonggui Series & \\
  \hline\noalign{\smallskip}
Qi            & 550,600 & 550,600 & 550,600 & 550,600 & 550,600 \\
Intercalation & 201,850 & 202,050 & 202,050 & 202,050 & 202,050 \\
Revolution    & 131,904 & 131,904 & 130,205 & 130,205 & 130,205 \\
Crossing   & 260,187.86 & 260,388 & 260,388 & 260,388 & 260,388 \\
  \noalign{\smallskip}\hline
\end{tabular}
\ec
\tablecomments{0.86\textwidth}{$^1$Values when the epoch is the winter solstice
of 1280.}
\end{table}

As can be seen in Table~\ref{tab1}, the values of interval constants 
in Datongli and Naepyeon are identical to each other. 
Interestingly, the numbers for the Shoushili of the Goryeosa
are the same as those for the Naepyeon and the Datongli, except for 
the Revolution interval constant.

\subsection{Modern calculations}
\label{modern}

In modern calculations, we use the astronomical algorithms of 
\cite{meeus1989,meeus1998}
and the DE406 ephemeris of \cite{standish1997}. 
In addition, we use Besselian elements to calculate the solar eclipse 
time in a local circumstance. Although \cite{muckes1983} tabulated 
Besselian elements for the solar eclipses ranging from $-$2003 to 2526, 
they presented only the first order coefficients in each element. Hence, 
we use Besselian elements extracted from DE406 ephemeris to increase 
our accuracy \citep[e.g.,][]{lee2008b}. One of the important parameters to calculate 
ancient astronomical phenomena is $\Delta T$, difference between the 
universal time (UT) and the dynamical time (TD). To estimate $\Delta T$ for a 
given year, we employ the cubic spline interpolation 
method \citep[see][]{press1992} using the data obtained recently by 
\cite{morrison2004}. To directly compare the results of modern 
calculations with those by the Shoushili-affiliated calendars, 
we convert the universal time into the local apparent solar time by 
correcting the equation of time. Lastly, we assume that the locations 
of Beijing, Nanjing, and Seoul are at 39$^{\circ}$ 55$^{\prime}$ N 
and 116$^{\circ}$ 25$^{\prime}$ E, 32$^{\circ}$ 3$^{\prime}$ N and
118$^{\circ}$ 53$^{\prime}$ E, and 37$^{\circ}$ 34$^{\prime}$ N and 
126$^{\circ}$ 59$^{\prime}$ E, respectively.

\section{Results}
\label{results}

\subsection{Values of interval constants}
\label{constants}

A winter solstice was used as the epoch 
in ancient Chinese calendars. In the Shoushili, the values 
of the interval constants were based on the winter solstice of 1280, 
as mentioned earlier. It is known that Shoujing Guo determined 
the date of the winter 
solstice from gnomon shadow measurements \citep{chen1983,li2005}. 
He used a tall gnomon and estimated the moment when the length of
the shadow, caused by the Sun, 
is the longest, based on several days' observations. In modern times, 
the winter solstice is calculated as the time when the Sun is passing 
the ecliptic longitude ($\lambda$) of 270$^{\circ}$, using an astronomical 
ephemeris such as DE406. For the details on how Shoujing Guo determined 
the times of the winter solstice and of the lunar perigee and descending 
node passages, refer to Yuanshi.

In Table~\ref{tab2}, we present the dates related to four interval 
constants in 
the Shoushili-affiliated calendars along with the results from modern 
calculations. All dates are in the Julian calendar, 
in units of the apparent solar time at Beijing unless otherwise mentioned. 
The difference in the equation of time according to the regions
(i.e., Beijing, Nanjing, and Seoul), is negligible. 
Hence, we can easily convert the times at Beijing into times at other 
regions by correcting only the longitudinal difference. Therefore, the time 
at Seoul, for example, is obtained by adding the time at Beijing with 
+42.26\,min (i.e., 10.566$^{\circ}$ of longitudinal difference between 
Beijing and Seoul).

\begin{table}
\bc
\begin{minipage}[]{150mm}
\caption[]{Summary of the dates in the Shoushili-affiliated calendars
and modern calculations relating to the four interval 
constants\label{tab2}.}\end{minipage}
\setlength{\tabcolsep}{2.5pt}
\small
 \begin{tabular}{lccccll}
  \hline\noalign{\smallskip}
Item     & \multicolumn{2}{c}{Shoushili-affiliated calendars (A)} & 
       Modern Calculation (B) & B $-$ A & \multicolumn{1}{c}{Calendar$^7$} &
Remark\\
\cline{2-4}
     & Julian Date & JD $-$ 2188925.56 & JD $-$ 2188925.56 & (min) &  \\
  \hline\noalign{\smallskip}
WS1280$^1$   &Dec. 14.060000 &~~~~~~0.000000 &~~~~~0.011638 & 16.8 & S, D, N &Epoch \\
\hline
MFDSC$^2$&Oct. 20.000000 &$-$55.060000 &$-$55.060000 &~~0.0  & S, D, N &Qi \\
\hline
MNM$^3$  &Nov. 23.875000 &$-$20.185000 & --          & --   & S   &
Intercalation \\
         &Nov. 23.855000 &$-$20.205000 & --          & --   & D, N \\
\hline
NM$^4$   &Nov. 24.211966 &$-$19.848034 &$-$19.840819 & 10.4 & S &
(Intercalation)      \\
         &Nov. 24.191856 &$-$19.868144 &$-$19.840819 & 39.3 & D, N    \\
\hline
LPP$^5$  &Nov. 30.869600 &$-$13.190400 &$-$13.355960 & 238.4& S &
Revolution      \\
         &Dec. 01.039500 &$-$13.020500 &$-$13.355960 & 483.1& D, N    \\
\hline
LDNP$^6$ &Nov. 18.041214 &$-$26.018786 &$-$25.863352 &$-$223.8& S & 
Crossing      \\
         &Nov. 18.021200 &$-$26.038800 &$-$25.863352 &$-$252.6& D, N    \\
  \noalign{\smallskip}\hline
\end{tabular}
\ec
\tablecomments{0.86\textwidth}{
$^1$Winter solstice of 1280, $^2$Midnight of the first day in a sexagenary 
cycle on counting backwards from the epoch, $^3$Mean new moon, $^4$new moon, 
$^5$Lunar perigee passage, $^6$Lunar descending node passage, 
$^7$S: Shoushili, D: Datongli, N: Naepyeon}
\end{table}

In the table, the first column contains items related to the 
interval constants; WS1280 is the winter solstice of 1280, i.e., the epoch of
the Shoushili-affiliated calendars. 
MFDSC is the midnight of the 
first day in the sexagenary cycle (i.e., Jiazi day) 
before the epoch. MNM and MN are mean new 
moon and new moon, respectively, and LPP and LDNP are lunar perigee 
and descending node passages, respectively. The second column contains the 
Julian dates derived from the epoch
and the values of interval constants in the Shoushili-affiliated calendars, 
except for WS1280, the epoch itself, and NM. 
The third column is the day number obtained by subtracting 
2188925.56\,d (i.e., the epoch) from the Julian day number (JD)
corresponding to the date 
given in the second column. The fourth and fifth columns are the results 
of modern calculations and the difference between the modern
calculations and the values derived from 
the Shoushili-affiliated calendars, respectively. The sixth column 
represent the calendar; S is Shoushili, D Datongli, 
and N Naepyeon. The last column contains the interval constants 
related with the items in the first column.

According to modern calculations, $\Delta T$ in 1280 is 532.6\,s and JD of 
the winter solstice of the year at Beijing is 2188925.571638\,d 
(i.e., 1280 December 14.071638), which is obtained
by correcting the equation of time by 
$-$0.087\,min. A difference of +16.8 min compared to modern calculation 
shows that Shoujing Guo accurately estimated 
the epoch in Shoushili (see Table~\ref{tab2}). 
Although the time difference between Shoushili and modern calculations
is 16.8\,min, there is no change in a date.
Hence, the MFDSCs in the Shoushili and modern calculations are the same
as the 56th day in the sexagenary cycle (i.e., Jiwei day), October 20.0.
Because of this,
the Qi interval constant, which is the length between the epoch and 
the MFDSC, has the same difference in the epoch,
i.e., 16.8\,min.
To verify our calculation,
we compute the winter solstice of 2010 using $\Delta T$ = 65.9\,s 
\citep{nao2009}, compare 
the result with the data of the \citeauthor{cas2009}
(\citeyear[shortly CAS2009]{cas2009}), and 
find a good agreement in the values, with the JD being 2455552.485037\,d 
(i.e. 2010 December 21.985037) in UT \cite[see als][]{kasi2009}. 

Although it is known that Shoujing Guo also determined the Intercalation 
interval constant based on observations, there is not much detail
on the calculations he used \citep{li1998a}. 
Hence, we calculated the date of the new moon 
in 1280 November as an indirect method to verify the Intercalation interval 
constant. The dates are calculated to be JD 2188905.711966 and 
JD 2188905.719181\,d (or 1280 November 21.344808 in UT) by Shoushili 
and modern calculations, respectively, which gives just a difference 
of only +10.4 min. To check our calculation, we also compare the time of the 
new moon with the data provided by NASA\footnote{http://eclipse.gsfc.nasa.gov/
phase/phasecat.html} and find that the difference is less than 1\,min. 
The exact times at which the astronomers of the Yuan dynasty 
modified the values of the interval constants of the Shoushili 
are not known. On using
Datongli's value of the Intercalation interval constant, 
the difference increases, becoming $\sim$39.3\,min, at least for the new moon 
time on 1280 November. We discuss the values for other periods in 
the next subsection.

Unlike the Qi and Intercalation interval constants, 
we find relatively large differences in the remaining constants. 
According to modern calculations, the lunar perigee and the descending node 
passage times at that time are JD 2188912.204040 and JD 2188899.696648\,d, 
respectively. That is, Revolution and Crossing 
interval constants show the difference of $\sim$238.4 (cf. $\sim$216\,min; 
\citeauthor{chen2006} \citeyear{chen2006}) 
and $\sim$223.8\,min, respectively, compared with modern calculations.
In particular, there is a large change in the Revolution interval constant
in the Datongli, i.e., +1699 Parts ($\sim$244.7\,min). Hence, the difference 
is also larger in proportion to the 
amount, i.e., $\sim$483.1 (= 238.4 + 244.7)\,min in Datongli
(refer to Tables~\ref{tab1} and \ref{tab2}).

\subsection{New moon time}
\label{nmt}

In the Shoushili-affiliated calendars, the time of any phase of the Moon, 
for example, new moon, is determined in the following manners. 
First calculate the mean new moon time using the Intercalation interval 
constants and the length of the synodic month 
(i.e., 295305.93 Parts in the Shoushili-affiliated calendars). The new 
moon time is then determined by using the Revolution interval constant 
and by considering 
the motions of the Sun and the Moon. 
Because there is no Li-Cheng in Yuanshi, 
we use the values of Shoushili-Li-Cheng preserved in Kyuganggak for the 
the motions of the Sun and the Moon. For the sake of completeness, we check
the values of Shoushili-Li-Cheng against Mingshi 
and Naepyeon and find that all values are identical to each other, 
except for a few typos in each book.

Comparing the length of the synodic month in the Shoushili with that 
obtained by modern calculations (i.e., 29.530587\,d in 1280;
refer to CAS2009), 
the difference is less than 1\,s. Hence, we ignore the error in the 
length of the synodic month in the Shoushili. 
Instead, it is worth noting that 
the motion of the Sun speeds up or slow down before and after the perihelion 
or aphelion and not the winter or summer solstice as mentioned in the 
Shoushili. 
However, it is well known that the time of the winter solstice was very 
close to the perihelion passage time of the Earth around 1280. According to 
our investigations, the Earth passed the perihelion on the
JD 2188925.078000\,d resulting in the difference of 0.48200\,d compared to
the time of the winter solstice in 1280.

\begin{figure}
   \centering
   \includegraphics[width=7.0cm, angle=0]{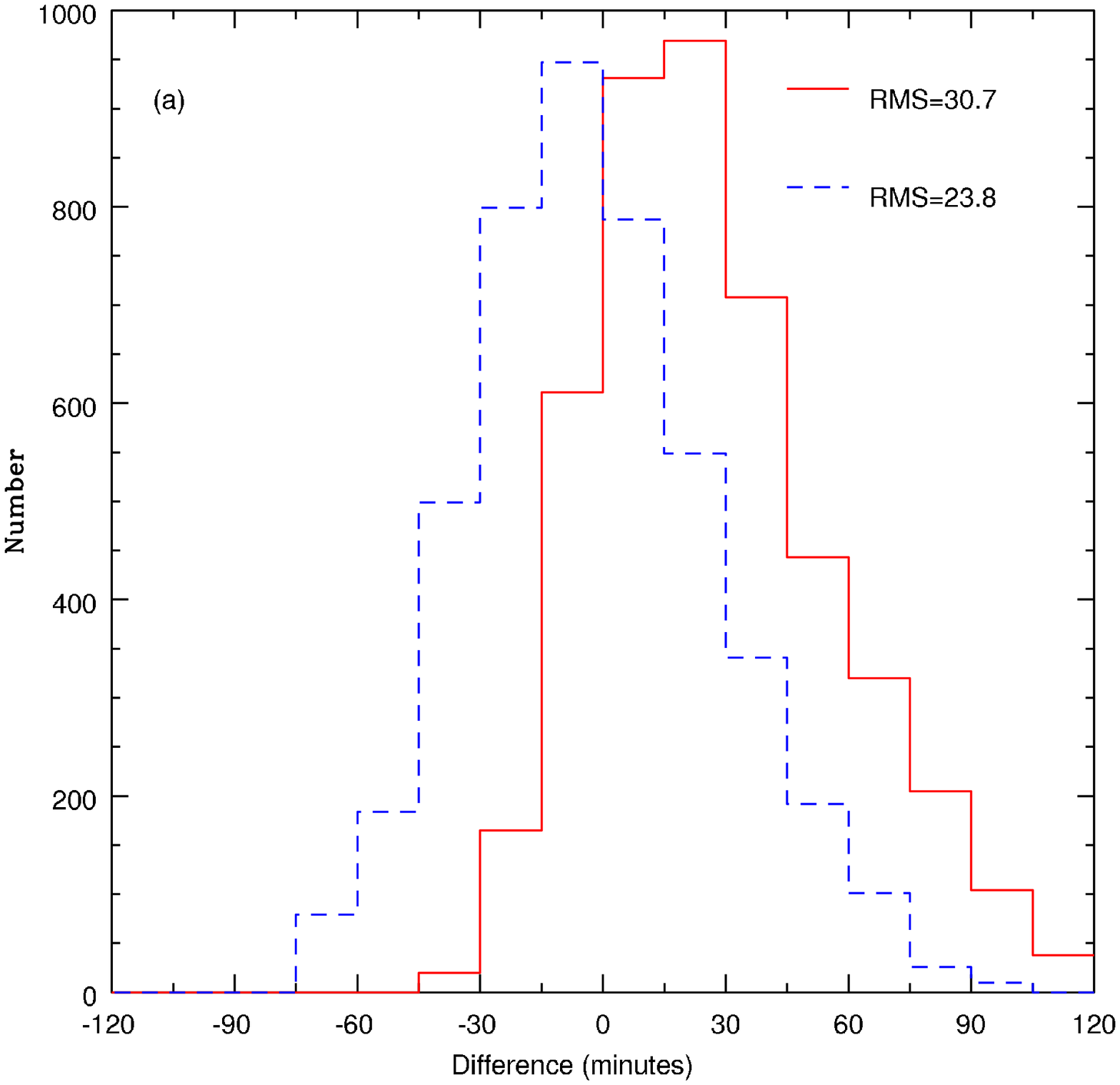}
   \includegraphics[width=7.0cm, angle=0]{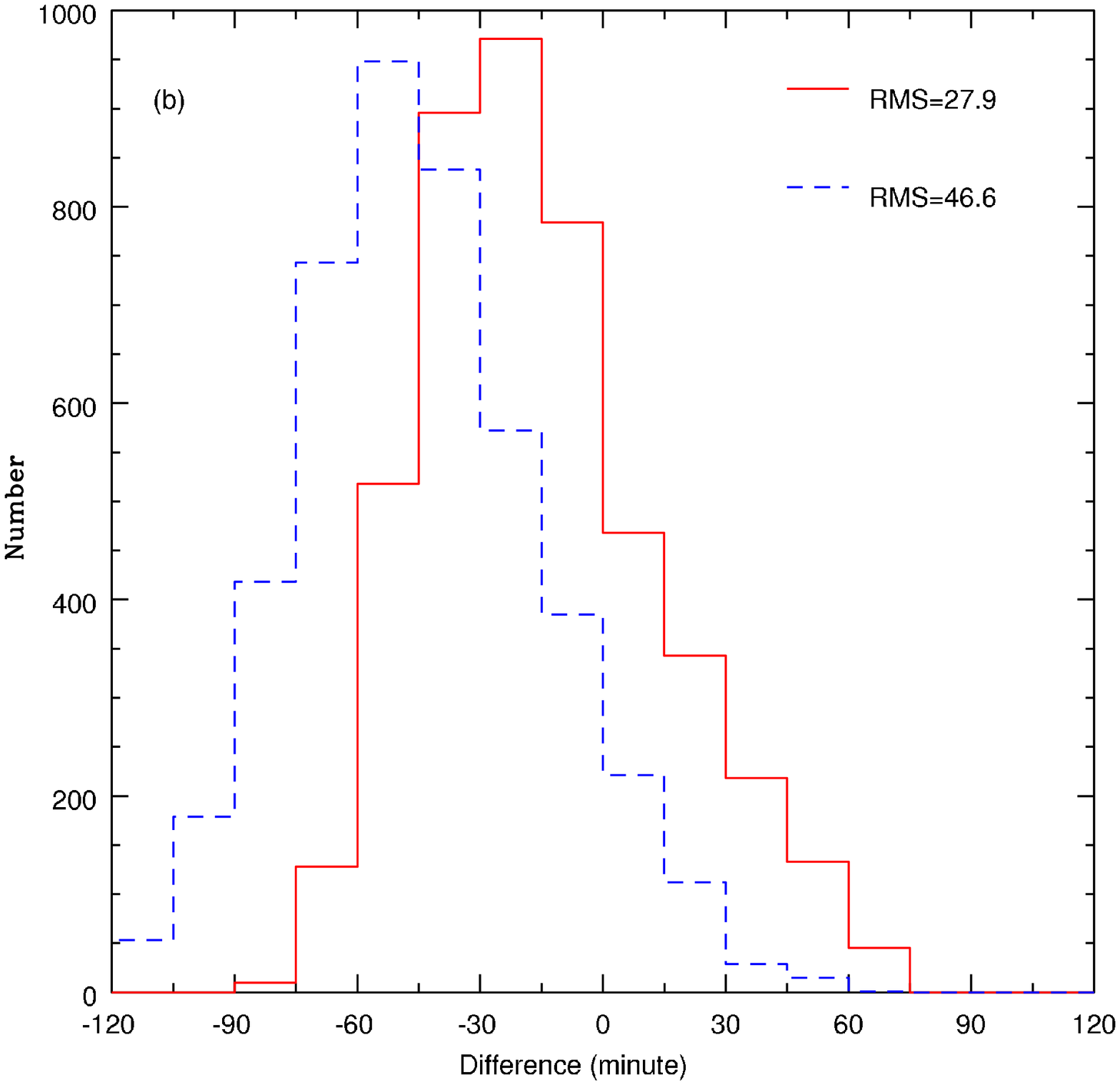}
   \caption{
The number distribution showing the differences in the new moon times between 
the Shoushili-affiliated calendars and modern calculations at 
(a) Beijing and (b) Seoul. 
The red-solid and blue-dotted lines represent the 
results for Shoushili and Datongli (Naepyeon in (b)), respectively.}
   \label{Fig1}
   \end{figure}

To evaluate the effect of the Intercalation and the Revolution interval 
constants 
in determining the new moon times, we compute those times for the 
Shoushili-affiliated calendars for the years ranging from 1280 to 1644, 
compare them with the results of modern calculations, and present the number 
distribution of the differences in Figure~\ref{Fig1} along with the 
root-mean-square (RMS) values. In the figure, the panels (a) and (b) 
show the results at Beijing and Seoul, respectively. 
The red-solid and blue-dotted lines represent the results for
Shoushili and Datongli (Naepyeon in Fig. 1(b)), respectively. 
In each panel, the horizontal and vertical axes represent the difference 
in units of minutes with the intervals of 15\,min and the number, 
respectively.

As shown in Figure~\ref{Fig1}, the new moon times given by the Datongli are 
on an average more accurate than those given by Shoushili at Beijing, i.e., 
RMSs are $\sim$23.8\,min 
(cf. $\sim$21\,min; \citet{li1998a,li1998b}) 
and $\sim$30.7\,min, 
respectively.  Therefore, it can be said that the Datongli (or the later
Shoushili) is one of 
the better calendars in China. However, Shoushili gives 
better results than Datongli in Korea, That is, the RMSs by the former
and latter calendars are 
$\sim$27.9 and $\sim$46.6\,min, respectively, in Korea. 
Interestingly, the RMS by Shoushili in Korea 
(i.e., RMS of $\sim$27.9 min) is even less than that in 
China (RMS of $\sim$30.7 min).

\subsection{Sunrise and sunset time}
\label{srsstime}

\begin{figure}
   \centering
   \includegraphics[width=7.0cm, angle=0]{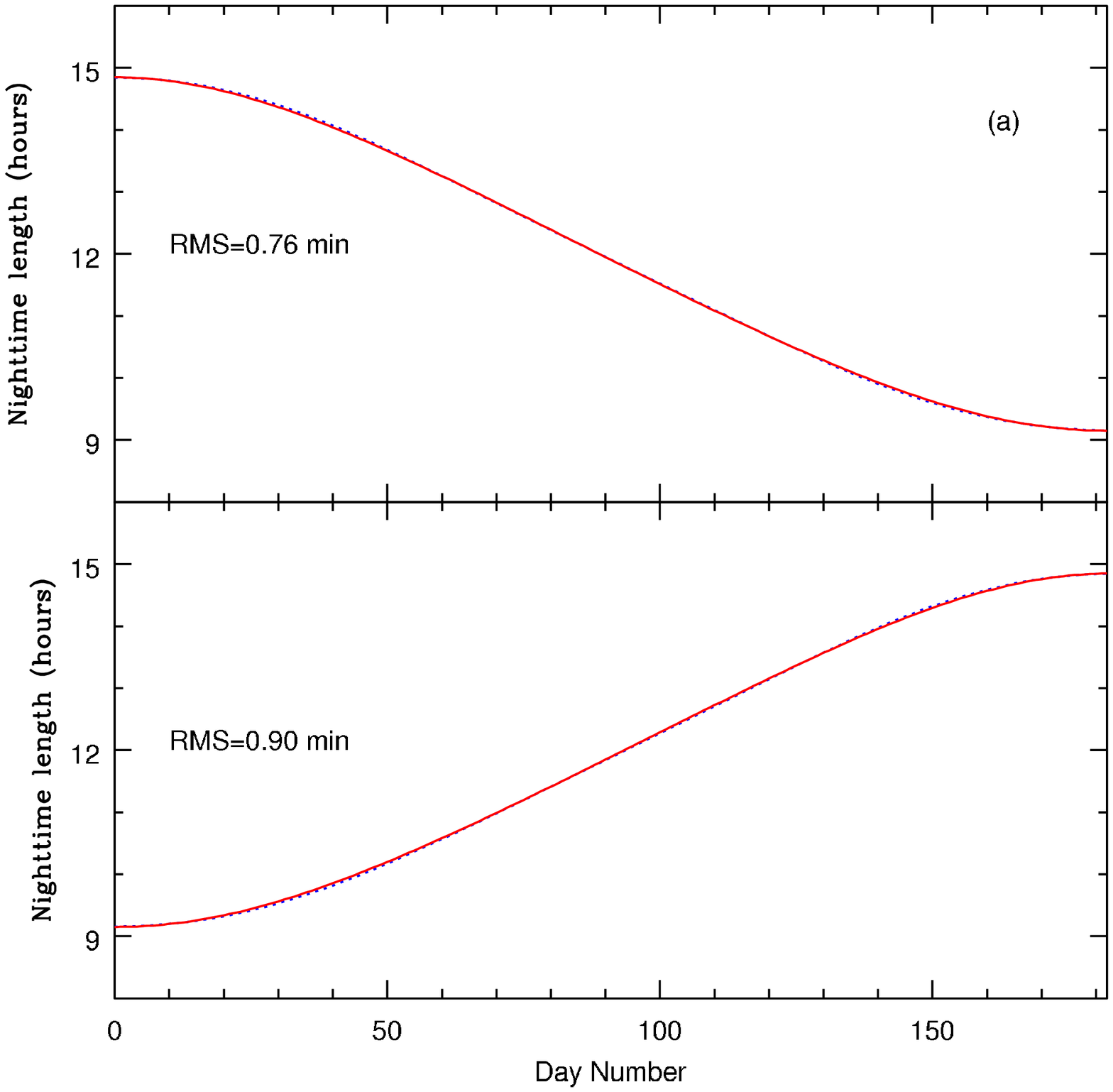}
   \includegraphics[width=7.0cm, angle=0]{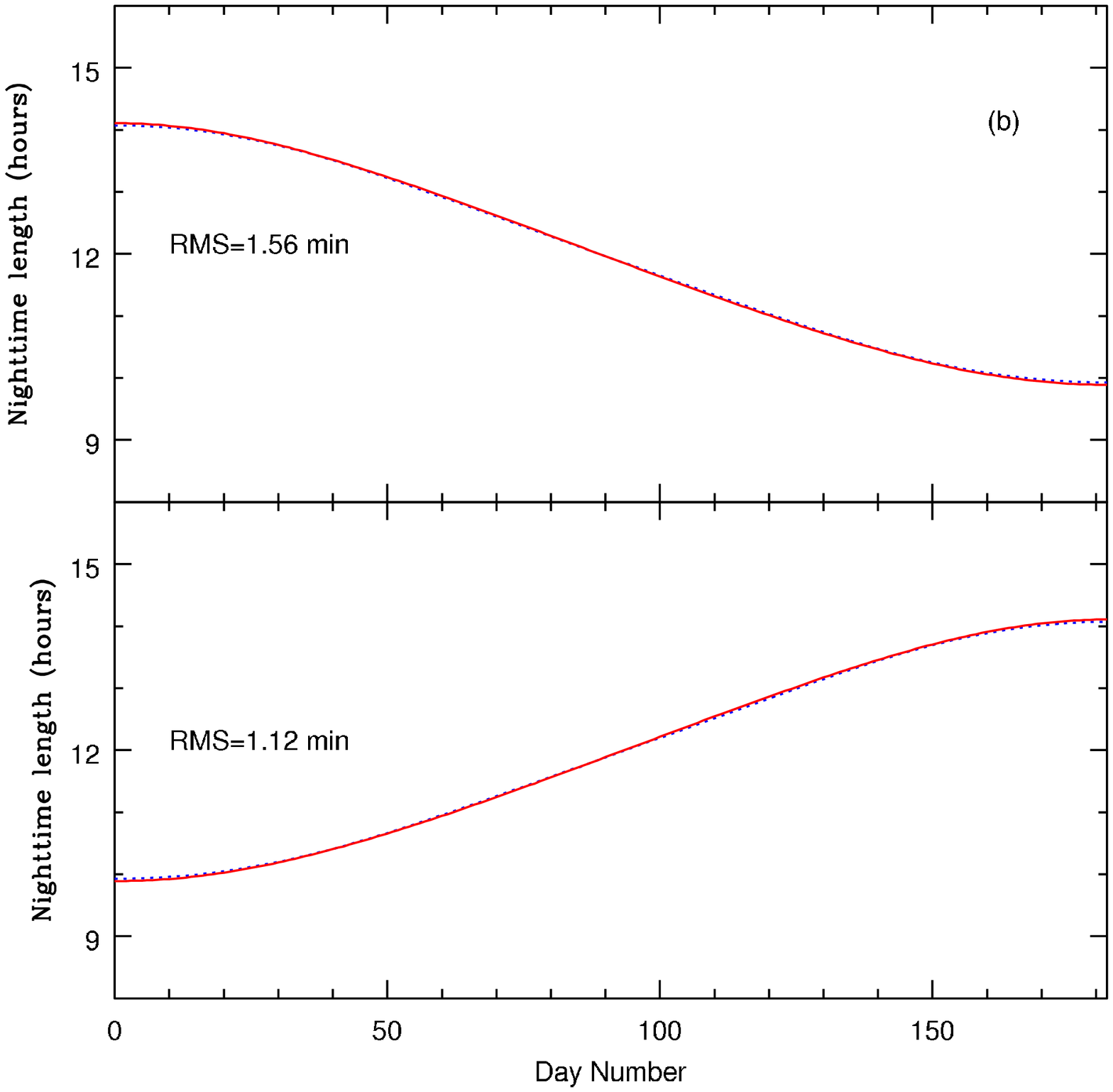}
   \caption{
Nighttime lengths (a) from Shoushili-Li-Cheng at Beijing and 
(b) from Mingshi (Taiyin-Tonggui) at Nanjing. 
The blue-dotted and red-solid lines indicate
results from the literature and the modern calculations,
respectively, at each region. 
In each panel, the upper and bottom are nighttime lengths after the winter 
and summer solstices, respectively. For more details, refer to text.}
   \label{Fig2}
   \end{figure}

One way to assess the overall accuracy of the four interval constants
is to check the solar eclipse time. Before that, we verify the timings of
sunrise (SR) and sunset (SS) presented in the calendar books because 
these times are used for calculating the times of the solar eclipses. 
In Figure~\ref{Fig2},
we depict the nighttime lengths (i.e., period from SS to SR) 
from the Shoushili-Li-Cheng and the Mingshi along with 
the results of modern calculations. In the figure, the blue-dotted lines 
represent the results 
(a) from Shoushili-Li-Cheng at Beijing and (b) from Mingshi
at Nanjing, while the red-solid lines are the results from modern 
calculations at each region. In each panel, the upper and bottom panels show 
the nighttime lengths after winter and summer solstices, respectively. 
The horizontal and vertical axes represent the day number
and the nighttime length in units of hours, respectively.
According the Shoushili-affiliated calendars, the summer 
solstice of 1281 is June 14, i.e., JD = 2189107.5\,d 
\citep{lee2010}. In this study, we define 
the SR/SS time as the zenith distance ($z$) of 90$^{\circ}$, 
which is different from the modern definition of 
$z$ = 90$^{\circ}$ 50$^{\prime}$. 
The result for Naepyeon at Seoul is given 
in the work of \cite{lee2011}, and shows
a difference of less than 1\,min on average.

A table on the sunrise and sunset times is also contained in the 
Taiyin-Tonggui (Comprehensive Guide on the Sun), 
which is one of Tonggui series. According to our observations, 
there were some disagreements between Mingshi 
and Taiyin-Tonggui in the SR and SS times. 
We can easily figure out which document is incorrect by
checking Daybreak (period from midnight to sunrise
or from sunset to midnight) and
Dusk (period from midnight to sunset) 
Parts because the sum of both parts should be 10,000 Parts, i.e., one day. 
In spite of this check, there were two discrepancies. However, these can be 
ignored because the differences were less than 1 Part. Hence, we can 
consider that the values of the SR and SS times are identical in both 
documents. Although there is no statement in the Taiyin-Tonggui,
the Mingshi explicitly states that the SR and SS times are 
those at Nanjing. As shown in Figure~\ref{Fig2}, 
the RMSs are less than $\sim$1\,min, compared to modern calculations,
making it hard to distinguish from each other. 
Therefore, we can confirm that the SR and SS 
times given in the Shoushili-Li-Cheng are those at Beijing 
and that the statement on the SR and SS times in Mingshi is true.

\subsection{Solar eclipse time}
\label{solar}

Based on the four interval constants and the SR/SS times discussed above, 
we calculate the timings of the solar eclipse maximum 
according to the calendars 
and compare these with the results of modern computations. 
Prior to the comparison, we validate our calculations 
by the Shoushili-affiliated 
calendars using the records in the historical literature. In the Garyeong, 
the procedure for calculating solar and lunar eclipses by Naepyeon 
is described in great detail. A calendar book entitled Jiaosi-Tonggui
(Comprehensive Guide on the Eclipse), which is 
preserved in Kyuganggak, also lists step-by-step values
in the process of calculating 
several eclipses by Datongli. We find that the results 
of our calculations of Naepyeon and Datongli show exact agreement with those 
of the documents. In particular, we find that the solar eclipse times 
of Jiaosi-Tonggui come from the result that used the SR/SS times at Nanjing
and  not Beijing. 
We compare the 
calculations in the Shoushili with the records 
in the Mingshi. In the history book, the times for a total of 32 solar 
eclipses, calculated by Shoushili, 
are recorded \citep[see also][]{yabuuchi2006}. 
We find that all times match well 
except for four records: 707 June 1, 1059 January 1, 1061 June 1, and 
1162 January 1 in the luni-solar calendar. According to our computations 
based on the Shoushili, there is no solar eclipse in 707, whereas
the others show a difference of 1 Mark 
($\sim$15 min). However, we think that the actual differences 
would be smaller than 1 Mark, which is the significant digit in the records. 
For the hour systems used in ancient China and Korea, 
\citep[refer to][]{saito1995,lee2011}. 

\begin{figure}
   \centering
   \includegraphics[width=7.0cm, angle=0]{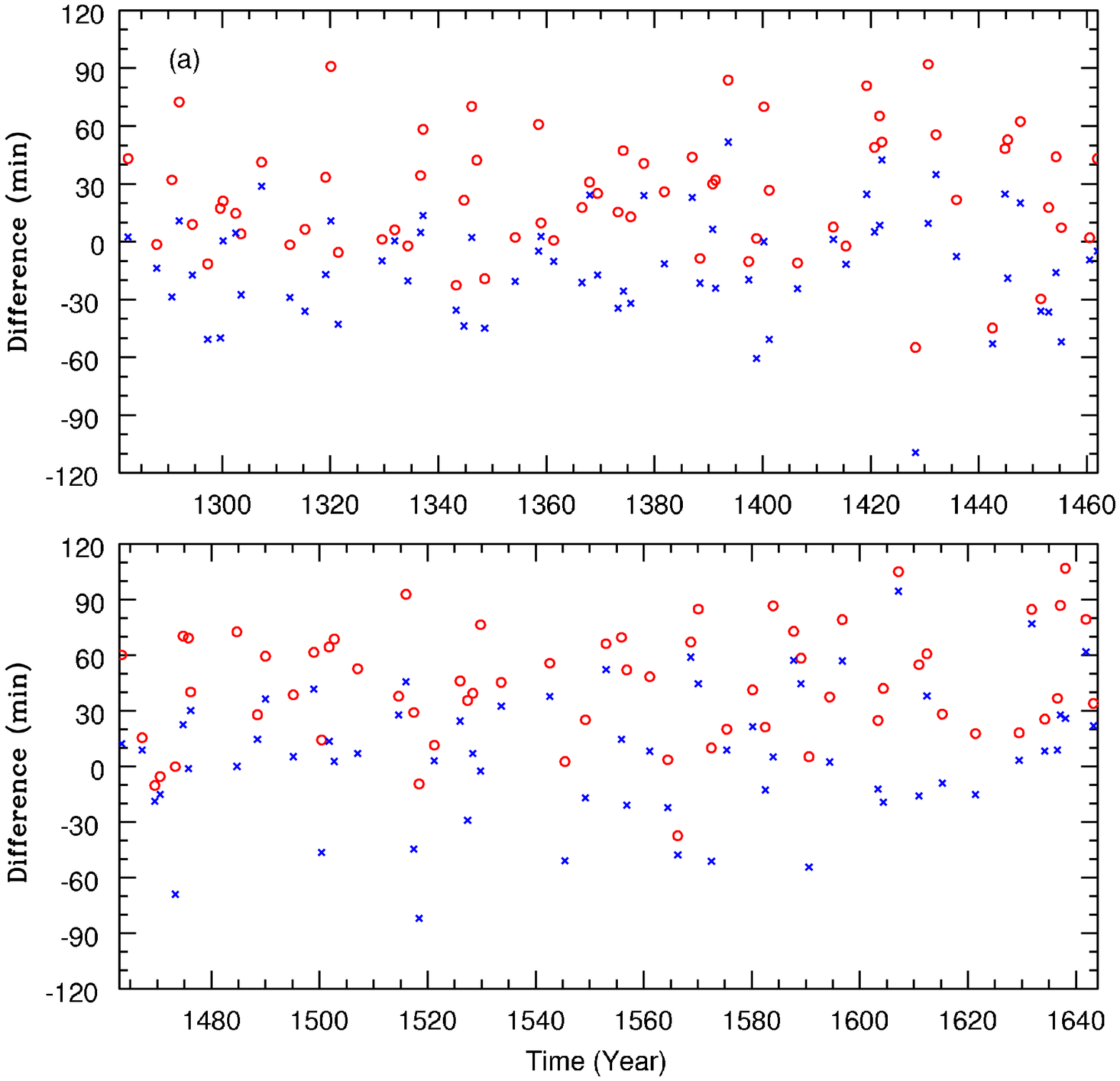}
   \includegraphics[width=7.0cm, angle=0]{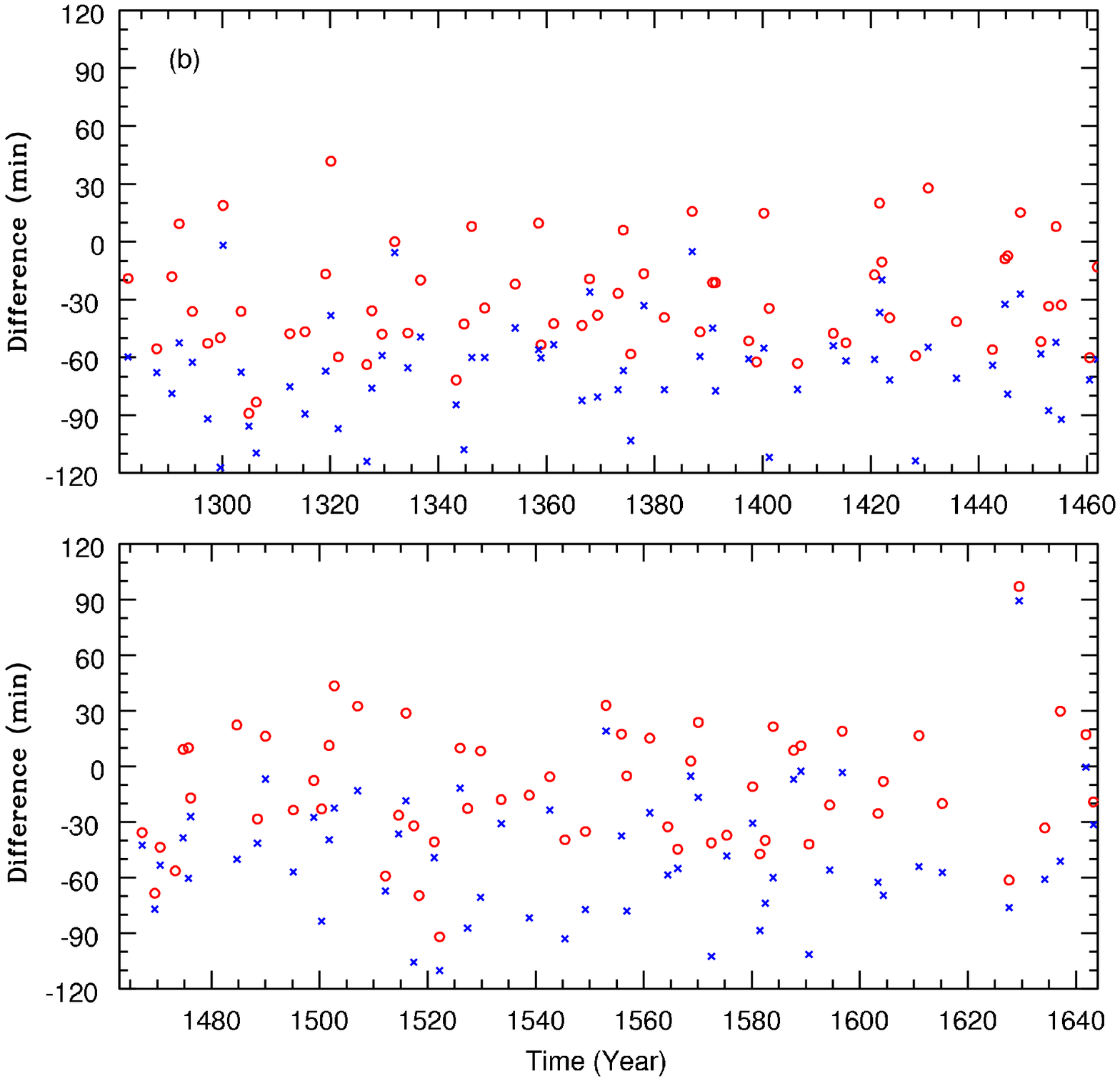}
   \caption{Differences in the solar eclipse maximum time between 
the Shoushili-affiliated calendars and modern calculations 
according to the region (a) at Beiing and (b) at Seoul. 
The circles and crosses represent the results for Shoushili and 
Datongli (Naepyeon in (b)), respectively.}
   \label{Fig3}
   \end{figure}

In this study, we restrict ourselves to the cases where the eclipse maximum
occurred during daytime, i.e., the Sun's altitude is greater than 
zero at the eclipse maximum. Figure~\ref{Fig3} shows the difference between 
the times given by the Shoushili-affiliated calendars (Ts) and by modern 
calculations (Tm), i.e., Ts $-$ Tm, between 1281 and 1644, along with
RMS. We use 
SR/SS times (a) at Beijing and (b) at Seoul. In each panel,
the circles and crosses represent the results for Shoushili and 
Datongli (Naepyeon in (b)), respectively.

According to this study, the RMSs for Shoushili and Datongli 
at Beijing are about $\sim$38.4 and $\sim$25.8 min, respectively. 
Meanwhile, \citet{li1997} also studied the difference between
the solar eclipse times given by Shoushili
(according to this study, the later Shoushili)
and by a modern ephemeris, and found that the 
RMS is $\sim$24\,min.
From the figure, 
we can easily see that the interval constants 
of the Datongli are better than those of Shoushili at Beijing, which is
similar to the case of the new moon times. The situation 
is reversed at Seoul (RMSs for Shoushili and Naepyeon are $\sim$50.8 and 
$\sim$77.1\,min, respectively). Thus, the interval constants of Shoushili 
give better results than those of Naepyeon for the prediction of the 
solar eclipse time in Korea.

\section{Summary}
\label{summary}

It is known that Shoushili, of the Yuan dynasty, is one of the most 
accurate calendars in the history of China. The court of the next 
dynasty, i.e., the Ming dynasty, revised the calendar and titled it
as Datongli. 
An example of the revisions is that the annual precession was abandoned.
In the Joseon dynasty of Korea, both calendars were referred to for the 
compilation of Naepyeon by the Joseon royal astronomers. 
With regard to the interval constants, 
the Joseon court adopted the values of Datongli (i.e., the
later Shoushili) in Naepyeon.
Although there are some differences,
particularly in the values of the 
interval constants, the Datongli and the Naepyeon are basically 
identical to the (early) Shoushili.

In this paper, we study the four interval constants given in
the Shoushili-affiliated calendars, which are related to the motions 
of the Sun and the Moon: the Qi, Intercalation, Revolution, and Crossing 
interval constants. We first compared the values of those interval 
constants with the results of modern calculations, and then investigate 
on the accuracy of the timings of the new moon and the solar eclipse 
maximum given by the 
Shoushili-affiliated calendars using the interval constants values adopted in 
each calendar, along with timings of sunrise and sunset. 
The following is the summary of our findings.

(1) In the Shoushili, the Qi and Intercalation interval constants are well 
determined (errors of $\sim$16.8 and $\sim$10.4\,min, respectively) 
while Revolution and Crossing ones are relatively poorly measured 
(errors of $\sim$238.4 and $\sim$223.8\,min, respectively). 
The latter two interval constants are worse in 
Datongli (errors of $\sim$483.1 and $\sim$252.6\,min, respectively).

(2) On calculating the new moon times for the period from 1280 to 1644
using Intercalation 
and Revolution interval constants of the Shoushili-affiliated calendars, 
the results show that the RMSs by Shoushili and Datongli are 30.7 and 
23.8\,min, respectively, at Beijing, and 27.9 and 46.6\,min, respectively, 
at Seoul. Therefore, it can be evaluated that Datongli is the better 
calendar in China but not in Korea. Moreover, Shoushili is, in general, 
more suitable for use in Korea rather than in China
in terms of the new moon time, at least for the period from 1280 to 1644.

(3) Unlike the interval constants, the timings of sunrise and sunset are 
accurately determined in each calendar book, with errors of less than 1\,min. 
In addition, the times listed in Shoushili-Li-Cheng are at Beijing,
as we confirmed, and in Mingshi are at Nanjing, as noted in the book.

(4) The timings of the solar eclipse maximum by Shoushili and Datongli ranging 
from 1281 to 1644 show RMSs of 38.4 and 25.8\,min, respectively, 
at Beijing, and 50.8 and 77.1\,min, respectively, at Seoul. 
Similarly to the new moon time, Shoushili gives better results than Datongli 
at Seoul in the calculation of the solar eclipse.

In the future, we think that more studies are needed to explain
why the RMSs for the timings for the new moon and the solar eclipse 
maximum are small, order of ten, compared to large errors in 
the Revolution and the Crossing interval constants of 
the Shoushili-affiliated calendars,
order of one hundred. One possibility might be the fact that
both constants have opposite 
signs in the differences when compared to modern calculations. 

\normalem
\begin{acknowledgements}
Ki-Won Lee is supported by Basic Science Research Program
through the National Research Foundation of Korea (NRF)
funded by the Ministry of Education (2013R1A1A2013747). 
\end{acknowledgements}
  
\bibliographystyle{raa}
\bibliography{shoushi}

\end{document}